\documentclass[aps,prc,twocolumn,superscriptaddress]{revtex4}

\usepackage{amsmath}  
\usepackage{amsfonts}
\usepackage{amssymb}
\usepackage{natbib} 
\usepackage{setspace}
\usepackage{graphicx} 
\usepackage{xspace}
\usepackage{cancel}
\usepackage{color}
\usepackage{float}
\usepackage{url}

\newcommand{\mi}{\mathrm{i}} 

\newcommand{\dfd}[3]{\hspace{-0.4em}\ensuremath{\frac{\mathrm{d}^{#1}#3}{(2\pi)^{#2}}}\,}

\newcommand{\eqn}[1]{Eq.~(\ref{#1})}
\newcommand{\fig}[1]{FIG.~\ref{#1}}
\newcommand{\tab}[1]{TABLE~\ref{#1}}

\def\Kbar{\overline{K}}
\def\K0bar{\overline{K^0}}

\def\Kbar{\overline{K}}

\newcommand{\be}{\begin{equation}}
\newcommand{\ee}{\end{equation}}
\newcommand{\bge}{\begin{equation}}
\newcommand{\ene}{\end{equation}}
\newcommand{\bea}{\begin{eqnarray}}
\newcommand{\eea}{\end{eqnarray}}
\newcommand{\bg}{\begin{eqnarray}}
\newcommand{\en}{\end{eqnarray}}

\begin{document}

\title{
  \vspace{-50mm}
  \begin{flushright}
    LFTC-17-1/1, ADP-17-23/T1029
  \end{flushright}
  \vspace{10mm}
  $\Phi$-meson--nucleus bound states}

\author{J.~J.~Cobos-Mart\'{\i}nez}
\affiliation{Laborat\'orio de F\'{\i}sica Te\'orica e Computacional-LFTC, 
Universidade Cruzeiro do Sul, 01506-000, S\~ao Paulo, SP, Brazil}
\affiliation{C\'atedra CONACyT, Departamento de F\'{\i}sica, Centro de Investigaci\'on
  y de Estudios Avanzados del Instituto Polit\'ecnico Nacional, Apartado Postal 14-740,
  07000, Ciudad de M\'exico, M\'exico}
\affiliation{Instituto de F\'{\i}sica y Matem\'aticas, Universidad Michoacana 
de San Nicol\'as de Hidalgo, Edificio C-3, Ciudad Universitaria, Morelia, 
Michoac\'an 58040, M\'exico.}
\author{K.~Tsushima}
\affiliation{Laborat\'orio de F\'{\i}sica Te\'orica e Computacional-LFTC, 
Universidade Cruzeiro do Sul, 01506-000, S\~ao Paulo, SP, Brazil}
\author{G.~Krein}
\affiliation{Instituto de F\'{\i}sica Te\'orica, Universidade Estadual 
Paulista, Rua Dr. Bento Teobaldo Ferraz, 271-Bloco II, 01140-070, S\~ao Paulo,
SP, Brazil}
\author{A.~W.~Thomas}
\affiliation{ARC Centre of Excellence for Particle Physics at the Terascale and CSSM, \\
Department of Physics, University of Adelaide,
Adelaide SA 5005, Australia}

\date{\today}

\begin{abstract}
  $\phi$-meson--nucleus bound state energies and absorption widths are calculated for seven selected
  nuclei by solving the Klein-Gordon equation with complex optical potentials. Essential input for the
  calculations, namely the medium-modified $K$ and $\Kbar$ meson masses, as well as the density
  distributions in nuclei, are obtained from the quark-meson coupling model. The attractive potential for
  the $\phi$-meson in the nuclear medium  originates from the in-medium enhanced $K\Kbar$ loop in
  the $\phi$-meson self-energy. The results suggest that the $\phi$-meson should form bound states 
with all the nuclei considered. However, the identification of the signal for these predicted bound states   
will need careful investigation because of their sizable absorption widths.
\end{abstract}


\maketitle

\date{\today}

\section{Introduction}

The properties of light vector mesons at finite baryon density, such as their masses and decay widths,  
have attracted considerable experimental and theoretical interest over the last few decades; see
Refs.~\cite{Hatsuda:1994pi,Leupold:2009kz,Hayano:2008vn} for recent reviews.
In part this has been related to their imputed potential  to carry information on the partial restoration
of chiral symmetry.
In 2007 the KEK-E325 collaboration reported a 3.4\% mass reduction of the
$\phi$-meson~\cite{Muto:2005za} and an in-medium decay width of $\approx 14.5$ MeV at normal
nuclear matter density.
These conclusions were based on the measurement of the invariant  mass spectra of $e^{+}e^{-}$ pairs
in 12 GeV p+A reactions, with copper and carbon being used as targets~\cite{Muto:2005za}.

Even though this result may indicate a signal for partial restoration of chiral symmetry in nuclear matter,
it is not possible to draw a definite conclusion solely from this. 
In fact, recently, a large in-medium $\phi$-meson decay width ($>$30 MeV) has been extracted at various
experimental facilities~\cite{Ishikawa:2004id,Mibe:2007aa,Qian:2009ab,Wood:2010ei,Polyanskiy:2010tj}, 
without observing any mass shift.
It is therefore evident that the search for evidence of a light vector meson mass shift in nuclear matter is
indeed a complicated issue and further experimental efforts are required in order to understand the
phenomenon better.
Indeed, the J-PARC E16 collaboration~\cite{JPARCE16Proposal,Csorgo:2014sat} intends to perform a more
systematic study for the mass shift of vector mesons with higher statistics than the above-mentioned
experiment at KEK-E325.

However, either complementary or alternative experimental methods are desired.
The study of the $\phi$-meson--nucleus bound states is complementary to the invariant mass
measurements, where only a small fraction of the produced $\phi$-mesons decay inside the nucleus and
may be expected to provide extra information on the $\phi$-meson properties at finite baryon density. 
Along these lines and motivated by the 3.4\% mass reduction reported by the KEK-E325 experiment, the
E29 collaboration at J-PARC has recently put forward a proposal~\cite{JPARCE29Proposal,
  JPARCE29ProposalAdd} to study the in-medium mass modification of the $\phi$-meson via the possible
formation of $\phi$-nucleus bound states~\cite{Buhler:2010zz,Ohnishi:2014xla} using the primary reaction
$\overline{p}p\rightarrow \phi\phi$.
Furthermore, there is also a proposal at the Thomas Jefferson National Accelerator Facility (JLab),
following the 12 GeV upgrade, to study the binding of $\phi$ (and $\eta$) to $^{4}$He~\cite{JLabphi}.
This new experimental approach~\cite{Buhler:2010zz,Ohnishi:2014xla,Csorgo:2014sat,JLabphi} for the
measurement of the $\phi$ meson mass shift in nuclei, will produce a slowly moving
$\phi$-meson~\cite{Buhler:2010zz,Ohnishi:2014xla,Csorgo:2014sat,JLabphi},  where the maximum nuclear
matter effect can be probed.
In this way, one may indeed anticipate the formation of a $\phi$-nucleus  bound state, where the
$\phi$-meson is trapped inside the nucleus.

Meson-nucleus systems bound by attractive strong interactions are very interesting objects; see
Refs.~\cite{Krein:2017usp, Metag:2017yuh} for recent reviews on the subject .
First, they are strongly interacting exotic many-body systems and to study them serves, for example, to
understand better the multiple-gluon exchange interactions, including QCD ``van der Waals''
forces~\cite{Appelquist:1978rt}, which are believed to play a role in the binding of the $J/\Psi$ and other
exotic heavy-quarkonia to matter (a nucleus)~\cite{Brodsky:1989jd,Luke:1992tm,Sibirtsev:1999jr,
  Gao:2000az, Beane:2014sda,Brambilla:2015rqa,Gao:2017hya,Kawama:2014pja,Iijima:2014cza,
Proceedings:2014rfa,Ohnishi:2014xla,Csorgo:2014sat}.
Second, they provide unique laboratories for the study of hadron properties at finite density, which may
not only lead to a deeper understanding of the strong interaction~\cite{Hatsuda:1994pi,Leupold:2009kz,
  Hayano:2008vn,Krein:2017usp, Metag:2017yuh} but of the structure of finite nuclei as
well~\cite{Stone:2016qmi,Guichon:2006er}.

A downward mass shift of the $\phi$-meson in a nucleus is directly connected with the possible
existence of an attractive potential between the $\phi$-meson and the nucleus, the strength of which is
expected to be of the same order as that of the mass shift.
Along these lines, various authors predict a downward shift of the in-medium $\phi$ meson mass
and a broadening of the decay width, many of them focusing on the self-energy of the $\phi$ meson
due to the kaon-antikaon loop. Ko {\it et al.}~\cite{Ko:1992tp} used a density-dependent kaon mass
determined from chiral perturbation theory and found that at normal nuclear matter density, $\rho_0$,
the $\phi$ mass decreases very little, by at most $2\%$, and the width  $\Gamma_\phi \approx 25$~MeV
and broadens drastically for large densities. Hatsuda and Lee calculated the in-medium  $\phi$ mass
based on the QCD sum rule approach~\cite{Hatsuda:1991ez,Hatsuda:1996xt},  and predicted a decrease
of 1.5\%-3\% at $\rho_0$.
Other investigations also predict a small downward mass shift and a large broadening  of the $\phi$ width
at $\rho_0$: Ref.~\cite{Klingl:1997tm} reports a negative mass shift of $ < 1\%$ and a decay width of 45
MeV; Ref.~\cite{Oset:2000eg} predicts a decay width of 22 MeV but does not report  a result on the mass
shift; and Ref.~\cite{Cabrera:2002hc} gives a rather small negative mass shift of
$\approx 0.81\%$ and a decay width of 30 MeV.
More recently, Ref.~\cite{Gubler:2015yna} reported a downward mass shift of $< 2\%$ and a large
broadening width of 45 MeV at $\rho_{0}$;  and finally, in Ref.~\cite{Cabrera:2016rnc},  extending the
work of Refs.~\cite{Oset:2000eg,Cabrera:2002hc}, the authors reported  a negative mass shift of
$3.4\%$ and a large decay width of 70 MeV at $\rho_0$.
The reason for these differences may lie in the different approaches  used to estimate the kaon-antikaon
loop contributions to the $\phi$-meson self-energy and this might have consequences for the formation of
$\phi$-meson--nucleus bound states.

From a practical point of view, the important  question is whether this attraction, if it exists, is sufficient
to bind the $\phi$ to a nucleus.
A simple argument can be given as follows. One knows that for an attractive spherical well of radius $R$
and depth $V_{0}$, the condition for the existence of a nonrelativistic $s$-wave bound state of  a particle
of mass $m$ is $V_{0}>\frac{\pi^2\hbar^2}{8mR^2}$.
Using $m = m_{\phi}^{*}$, where $m_{\phi}^{*}$ is the $\phi$-meson mass at normal nuclear matter density
found in Ref.~\cite{Muto:2005za} and $R= 5$ fm (the radius of a heavy nucleus), one obtains $V_{0}>2$ MeV.
Therefore, the prospects of capturing a $\phi$-meson seem quite favorable, provided that the
$\phi$-meson can be produced almost at rest in the nucleus.

An initial calculation of possible $\phi$-nucleus bound states was  carried out in
Ref.~\cite{YamagataSekihara:2010rb} for a few nuclei.
However, the theoretical potential on which this study was based~\cite{Cabrera:2002hc} was too weak, 
with only two bound states being found.
In order to remedy this,  the real part of the potential was scaled, without any theoretical basis, so as to
simulate a 3$\%$ mass reduction of the $\phi$-meson, that is, approximately equal to that reported in
Ref.~\cite{Muto:2005za}.
This (scaled) potential was mainly used to study the sensitivity of the formation spectra to the potential 
strength~\cite{Cabrera:2002hc}. Here, it was found that, as expected, whether or not the formation of
the $\phi$-meson bound state is possible  depends on the strength of the attractive potential 
between the $\phi$-meson and the nucleus.

In previous work~\cite{Cobos-Martinez:2017vtr} we studied the $\phi$-meson mass shift and decay width
in nuclear matter, based on an effective Lagrangian approach, by evaluating the $K\Kbar$ loop contribution 
in the $\phi$ self-energy, with the in-medium $K$ and $\Kbar$ masses  explicitly calculated by the
quark-meson coupling (QMC) model~\cite{Saito:2005rv}.
Here we extend our previous initial study~\cite{Cobos-Martinez:2017vtr}  to seven selected nuclei,
showing details of the calculated nuclear potential and computing the $\phi$-nucleus bound state
energies and absorption widths by solving the Klein-Gordon equation.
The nuclear density distributions for heavy nuclei studied (except for $^4$He), as well as the medium
modification of the $K$ and $\Kbar$ masses,  are explicitly calculated using the QMC
model~\cite{Saito:1996sf}.

This paper is organized as follows. In Sec.~\ref{sec:nuclmatt} we briefly discuss the computation and
present results for the mass shift and decay width of the $\phi$-meson in infinite (symmetric) nuclear
matter.
Using the results of Sec.~\ref{sec:nuclmatt}, together with the density profiles of the nuclei to be
studied, in Sec.~\ref{sec:finitenuclei} we present results for the real and imaginary parts of the
scalar $\phi$-nucleus potentials, as well as the corresponding bound state energies and absorption widths.
Finally, Sec.~\ref{sec:summary} is devoted to a summary and discussion.

\section{\label{sec:nuclmatt} $\phi$-meson self-energy in infinite nuclear 
matter}

The $\phi$-meson property modifications in nuclear matter, such as its mass  and decay width, are strongly
correlated to its coupling to the $K\Kbar$ channel, which is the dominant decay channel in vacuum. 
Therefore, one expects that a significant fraction of the density dependence of the $\phi$-meson
self-energy in  nuclear matter might arise from the in-medium modification  of the $K\Kbar$-loop 
in the $\phi$-self-energy intermediate state.

Here we use the effective Lagrangian approach of Ref.~\cite{Klingl:1996by} and briefly review the
computation of the $\phi$-meson self-energy in vacuum and in nuclear
matter made in Ref.~\cite{Cobos-Martinez:2017vtr}.
The interaction Lagrangian $\mathcal{L}_{\text{Int}}$ involves  $\phi K\Kbar$ and $\phi\phi K\Kbar$ couplings
dictated by a  local gauge symmetry principle:
\begin{equation}
 \label{eqn:Lint}
\mathcal{L}_{\text{Int}} = \mathcal{L}_{\phi K\Kbar} + \mathcal{L}_{\phi\phi K\Kbar},
\end{equation}
\noindent where
\begin{equation}
\label{eqn:phikk}
\mathcal{L}_{\phi K\Kbar} = \mi g_{\phi}\phi^{\mu}
\left[\Kbar(\partial_{\mu}K)-(\partial_{\mu}\Kbar)K\right],
\end{equation}
\noindent and
\begin{equation}
\label{eqn:phi2kk}
\mathcal{L}_{\phi\phi K\Kbar} = g^2_{\phi} \phi^\mu\phi_\mu \Kbar K.
\end{equation}
We use the convention for the isospin doublets:
\begin{equation}
\label{eqn:isospin}
K=\left(\begin{array}{c} K^{+} \\ K^{0} \end{array} \right),\;
\overline{K}=\left(K^{-}\;\overline{K}^{0}\;\right).
\end{equation}
%
\begin{figure}[t]
\centering
\includegraphics[scale=0.75]{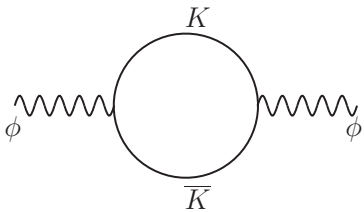}
\caption{\label{fig:phise} $K\Kbar$-loop contribution to the $\phi$ meson 
self-energy.}
\end{figure}
  
We note that the use of the effective interaction Lagrangian of \eqn{eqn:Lint} without the
term given in \eqn{eqn:phi2kk} may be considered as being motivated by the hidden gauge approach in
which there are no four-point vertices, such as \eqn{eqn:phi2kk}, that involve two pseudoscalar mesons
and two vector mesons~\cite{Lin:1999ve,Lee:1994wx}.
This is in contrast to the approach of using the minimal substitution to introduce vector mesons as gauge
particles where such four-point vertices do appear. However, these two methods have been shown to be
consistent if both the vector and axial vector mesons are
included~\cite{Yamawaki:1986zz,Meissner:1986tc,Meissner:1987ge, Saito:1987ba}.
Therefore, we present results with and without such an interaction.
We consider first the contribution from the  $\phi K\Kbar$ coupling given by Eq.~(\ref{eqn:phikk})
to the scalar part of the $\phi$ self-energy,  $\Pi_{\phi}(p)$, by evaluating the diagram of  \fig{fig:phise}.

For a $\phi$-meson at rest the scalar self-energy is given by
\begin{equation}
\label{eqn:phise}
\mi\Pi_{\phi}(p)=-\frac{8}{3}g_{\phi}^{2}\int\dfd{4}{4}{q}\vec{q}^{\,2}
D_{K}(q)D_{K}(q-p),
\end{equation}
\noindent where $D_{K}(q)=\left(q^{2}-m_{K}^{2}+\mi\epsilon\right)^{-1}$ is the kaon propagator;
$p=(p^{0}=m_{\phi},\vec{0})$ is the $\phi$-meson four-momentum  vector at rest, with $m_{\phi}$ the
$\phi$-meson mass; $m_{K} (=m_{\Kbar})$ is the kaon mass; and  $g_{\phi}$ is the coupling constant.

The integral in \eqn{eqn:phise} is divergent but it will be regulated using a phenomenological form factor,
with cutoff parameter $\Lambda_{K}$, as in Refs.~\cite{Krein:2010vp,Cobos-Martinez:2017vtr}. 
The sensitivity of the results to the cutoff value is analyzed below.

The coupling constant $g_{\phi}$ is determined in Ref.~\cite{Cobos-Martinez:2017vtr} from
the experimental value for the  $\phi \to K\overline{K}$ decay width in vacuum, corresponding to the
branching  ratio of $83.1\%$ of the total decay width (4.266 MeV)~\cite{PDG:2015}.
For the $\phi$ mass in vacuum, $m_{\phi}$, we use its experimental value $m_{\phi}^{\text{expt}}=1019.461$
MeV~\cite{PDG:2015}; while for the kaon mass $m_{K}$ we use the average of the experimental
values~\cite{PDG:2015} of the positive-charged and neutral kaons, $m_{K^{+}}^{\text{expt}}=493.677$ MeV
and $m_{K^{0}}^{\text{expt}}=497.611$, respectively.
We note that the effect of this tiny mass ambiguity on the in-medium kaon (antikaon) properties is
negligible. Then, we obtain the coupling $g_{\phi}=4.539$.
The $\Lambda_{K}$-dependent $\phi$-meson bare mass $m_{\phi}^{0}$ is fixed by fitting the physical
$\phi$-meson mass $m_{\phi}^{\text{expt}}$~\cite{Cobos-Martinez:2017vtr}, and the values obtained are
given in the caption to \tab{tab:phippties}.

\begin{figure}
\centering
\includegraphics[scale=0.23,angle=-90]{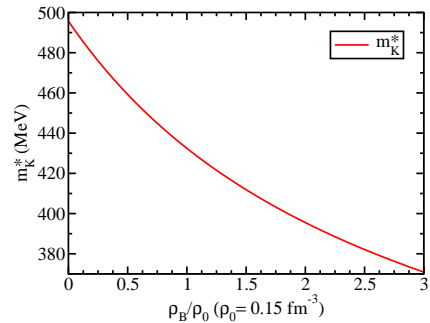}
\caption{\label{fig:mk} In-medium kaon (=antikaon) Lorentz scalar mass $m^*_{K}$.}
\end{figure}

The mass and decay width of the $\phi$-meson in vacuum ($m_{\phi}$ and $\Gamma_{\phi}$), as well as in
nuclear matter ($m_{\phi}^{*}$ and $\Gamma_{\phi}^{*}$),  are determined self-consistently in
Ref.~\cite{Cobos-Martinez:2017vtr} from 
\begin{eqnarray}
\label{eqn:phimass}
m_{\phi}^{2}&=&\left(m_{\phi}^{0}\right)^{2}+\Re\Pi_{\phi}(m_{\phi}^{2}), \\
\label{eqn:phidecay}
\Gamma_{\phi}&=&-\frac{1}{m_{\phi}}\Im\Pi_{\phi}(m_{\phi}^{2}).
\end{eqnarray}
The nuclear density dependence of the $\phi$-meson mass and decay width is driven by the
intermediate-state kaon and antikaon interactions with the nuclear medium. 
This effect enters through $m_{K}^{*}$ in the kaon propagators in~\eqn{eqn:phise}.
The in-medium mass, $m_{K}^{*}$, is calculated within the QMC model~\cite{Cobos-Martinez:2017vtr},
which has proven to be very successful in studying the properties of hadrons in nuclear matter and finite
nuclei. For a more complete discussion of the model see Refs.~\cite{Tsushima:1997df,Guichon:1989tx,
  Saito:2005rv}. Here we just make a few necessary comments. In order to calculate the in-medium
properties of $K$ and $\Kbar$, we consider infinitely large, uniformly symmetric, spin-isospin-saturated
nuclear matter in its rest frame, where all the scalar and vector mean field potentials, which are
responsible for the nuclear many-body interactions, become constant in the Hartree
approximation~\cite{Cobos-Martinez:2017vtr}.
In \fig{fig:mk} we present the resulting in-medium kaon Lorentz scalar mass (=antikaon Lorentz scalar mass), 
calculated using the QMC model, as a function of the baryon density. 
The kaon effective mass at normal nuclear matter
density $\rho_{0}=0.15$ fm$^{-3}$ has decreased by about 13\%. We also recall, in connection with the
calculation of the in-medium $K\Kbar$-loop contributions to the $\phi$-meson self-energy, that the
isoscalar-vector $\omega$ mean field potentials arise both for the kaon and antikaon.  However, they
have opposite signs and cancel each other. Equivalently, they can be eliminated  by a variable shift in the
loop calculation~\cite{Tsushima:1997df, Guichon:1989tx,Saito:2005rv} of the $\phi$ self-energy, and
therefore we do not show them here.

\begin{figure}[htb]
  \begin{tabular}{c}
    \includegraphics[scale=0.21,angle=-90]{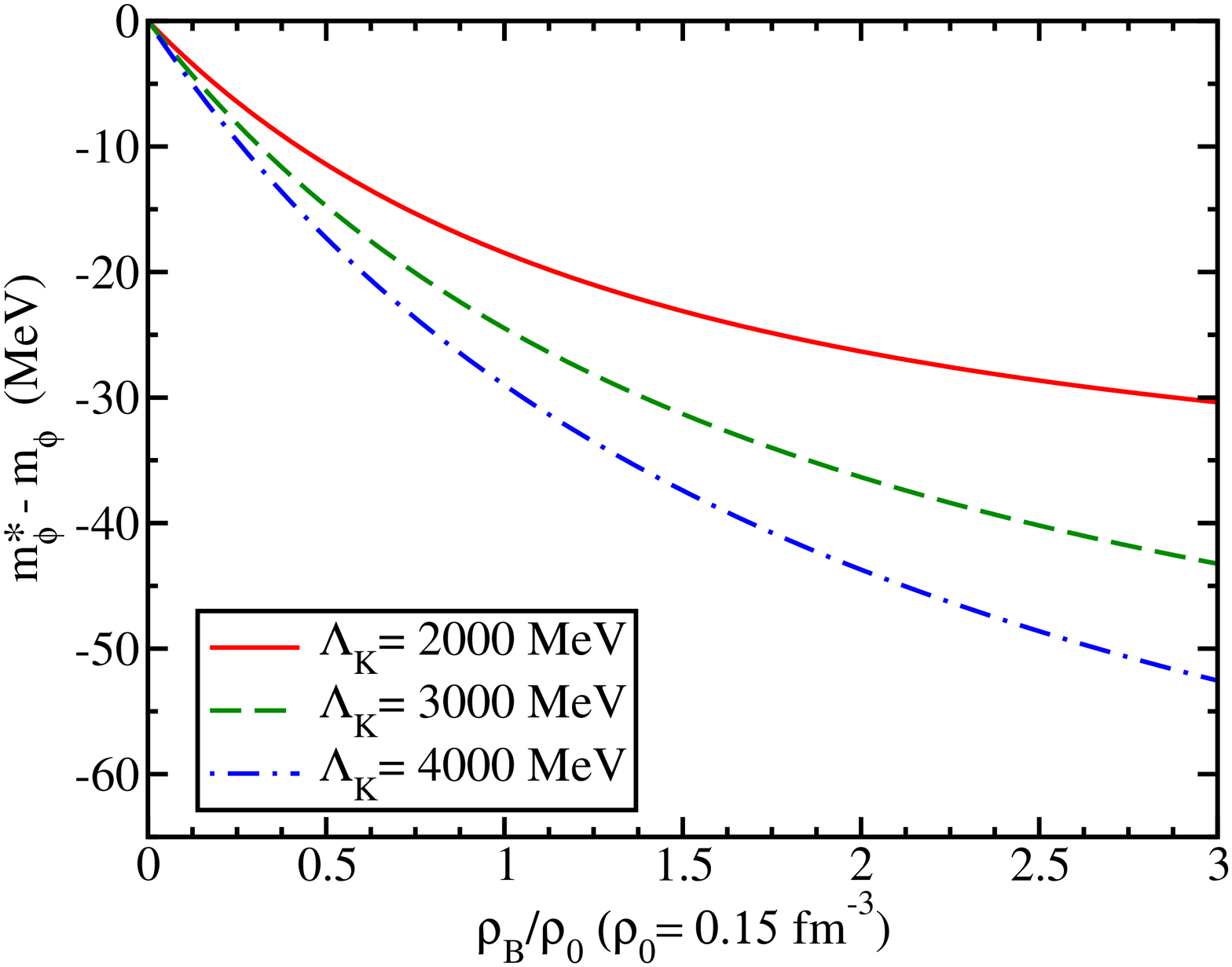} \\
    \includegraphics[scale=0.21,angle=-90]{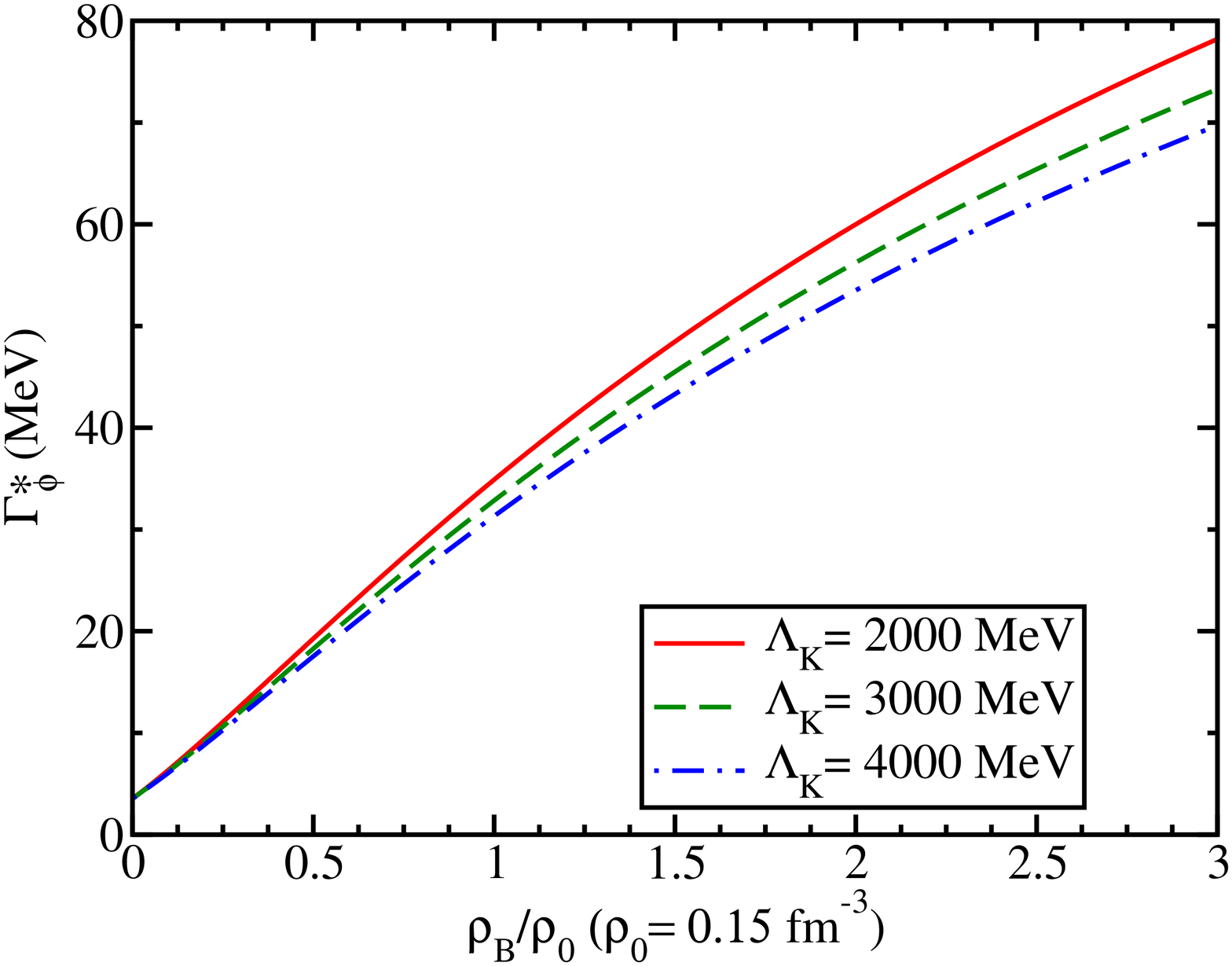}
  \end{tabular}
\caption{\label{fig:Xphi} $\phi$-meson mass shift (upper panel) 
and decay width (lower panel) in symmetric nuclear matter for three values 
of the cutoff parameter $\Lambda_{K}$.}
\end{figure}

In \fig{fig:Xphi}, we present the $\phi$-meson mass shift  (upper panel) and decay width (lower panel) as a
function of the nuclear matter density, $\rho_{B}$, for three values of the cutoff parameter $\Lambda_{K}$.
As can be seen, the effect of the in-medium kaon and antikaon mass change yields a negative mass shift
for the $\phi$-meson. This is because  the reduction in the kaon and antikaon masses enhances the
$K\Kbar$-loop contribution in nuclear matter relative to that in vacuum. For the largest value of the
nuclear matter density, the downward mass shift turns out to be a few percent at most for all values of
$\Lambda_{K}$. On the other hand, we see that $\Gamma_{\phi}^{*}$ is very sensitive to the change in the
kaon and antikaon masses: it increases rapidly with increasing  nuclear matter density, up to a factor of
$\sim 20$ enhancement for the largest value of $\rho_{B}$.  At normal nuclear matter density, $\rho_{0}$,
we see that the negative kaon and antikaon mass shift of 13\%~\cite{Cobos-Martinez:2017vtr} induces a
downward mass shift of the $\phi$-meson of just $\approx$ 2\%, while the broadening of the $\phi$-meson
decay width is an order of magnitude larger than its vacuum value.

\begin{figure}[htb]
\begin{tabular}{c}
\includegraphics[scale=0.23,angle=-90]{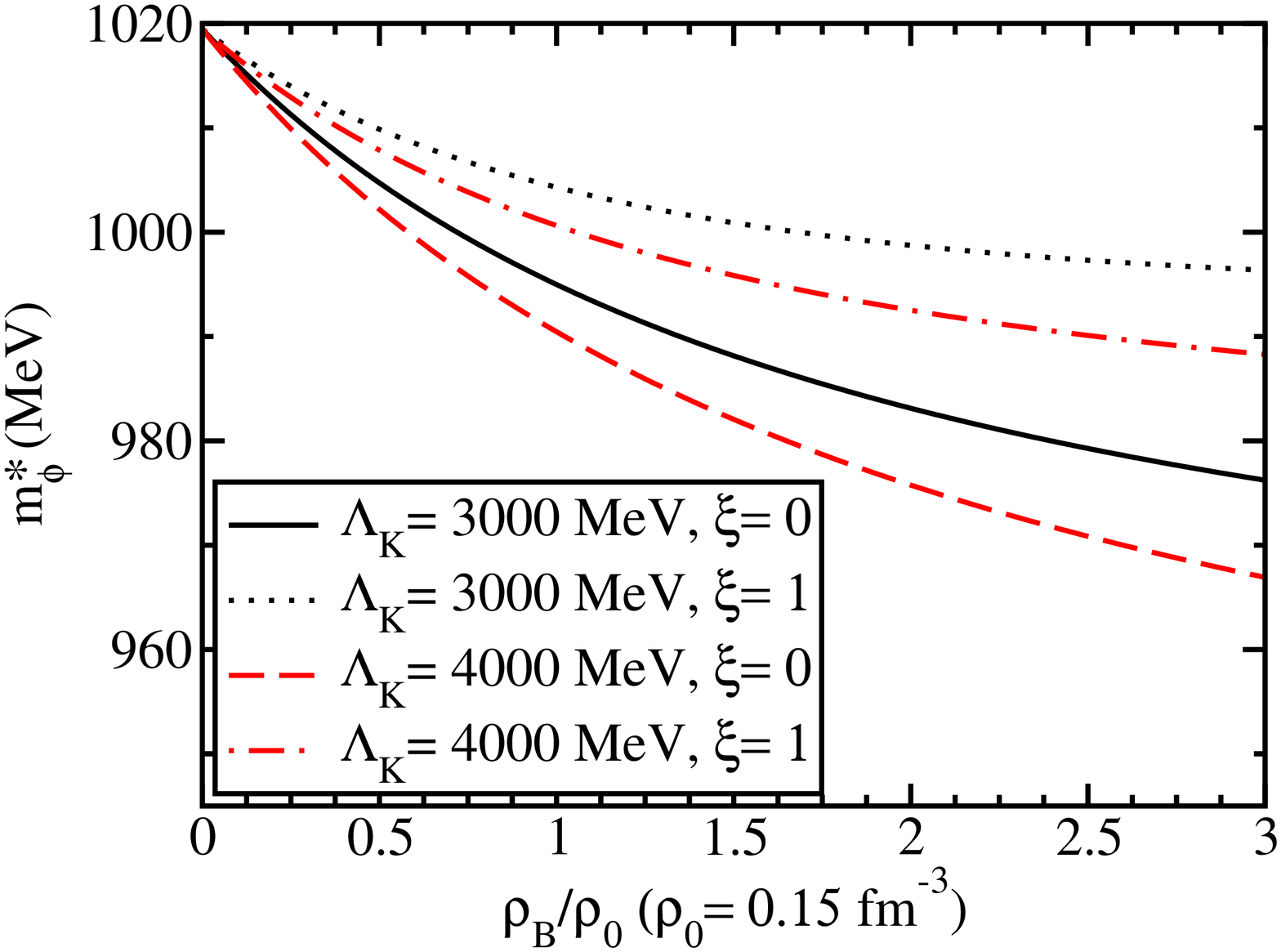} \\
\includegraphics[scale=0.23,angle=-90]{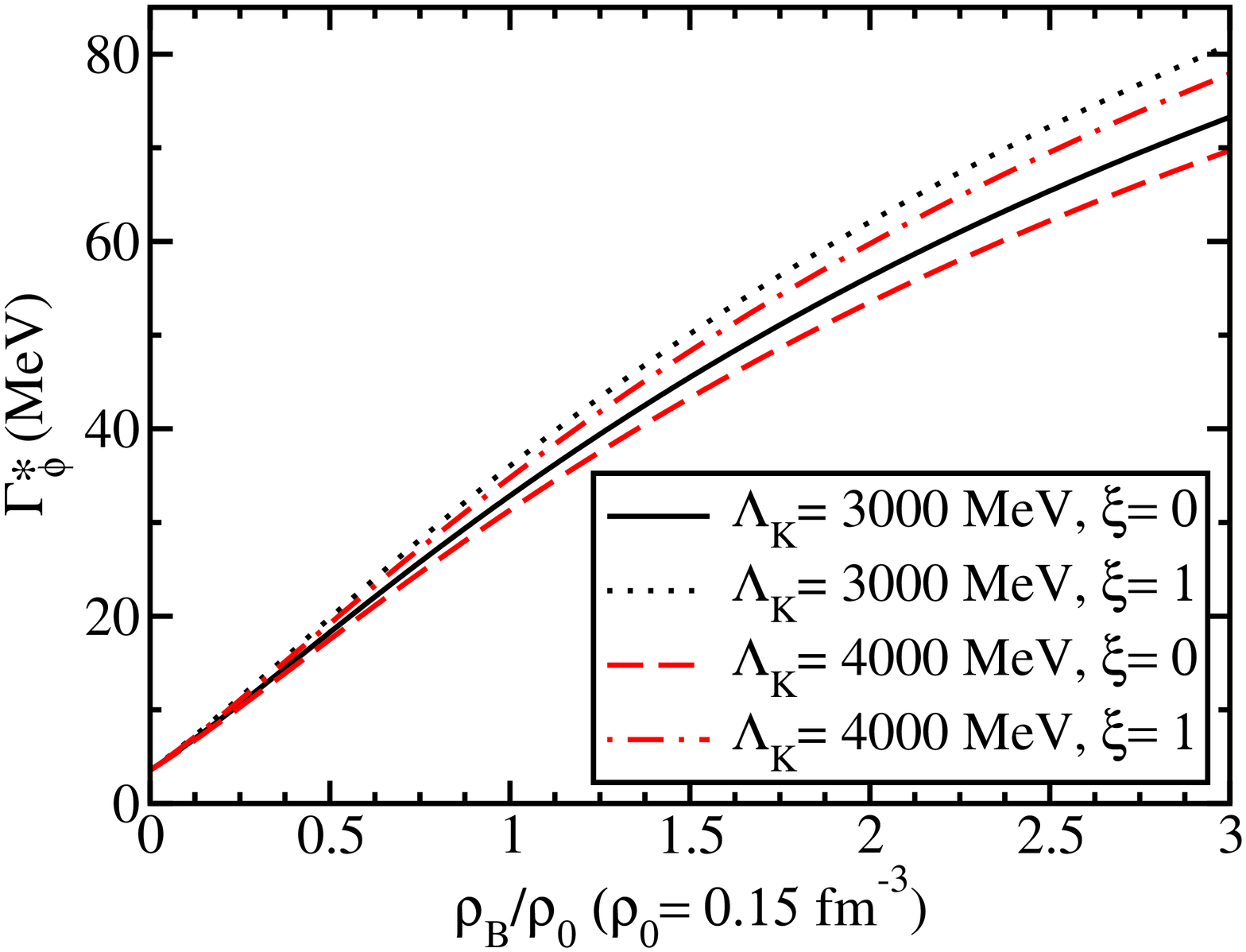}
  \end{tabular}
\caption{\label{fig:xiOn} Effect of adding ($\xi = 1$)  the $\phi\phi K \Kbar$ interaction of
  Eq.~(\ref{eqn:phi2kk}) on the in-medium $\phi$ mass (upper panel) and width (lower panel)
  for two values of the cutoff parameter $\Lambda_{K}$. }
\end{figure}

For completeness, and in connection with the paragraph just after \eqn{eqn:isospin}, we show in
\fig{fig:xiOn} the impact of adding the $\phi\phi K \Kbar$ interaction of Eq.~(\ref{eqn:phi2kk}) on the
in-medium $\phi$ mass and decay width.
We have used the notation that $\xi = 1 (0)$ means  that this interaction is (not) included in the calculation
of the  $\phi$ self-energy. 
As can be seen, one still gets a downward shift  of the in-medium $\phi$ mass when $\xi = 1$ as well as a
significant broadening of the decay width. In both cases, though, the absolute values are slightly different
from the $\xi = 0$ case.
In \tab{tab:phippties} we present the values for $m_{\phi}^{*}$ and $\Gamma_{\phi}^{*}$  at normal nuclear
matter density $\rho_{0}$ with and without the gauged Lagrangian of \eqn{eqn:phi2kk}.
In both cases, the effect of adding Eq.~(\ref{eqn:phi2kk}) can be compensated by the use of a larger
cutoff $\Lambda_{K}$ (compare the first and the last columns of \tab{tab:phippties}).

\begin{table}[h]
\begin{center}
\begin{tabular}{l|rrr}
\hline \hline
& $\Lambda_{K}= 2000$ & $\Lambda_{K}= 3000$  & $\Lambda_{K}= 4000$ \\
\hline
$m_{\phi}^{*}$ & 1000.9 (1009.5) & 994.9 (1004.3) & 990.4 (1000.6) \\
$\Gamma_{\phi}^{*}$ & 34.8 (37.8) & 32.8 (36.0) & 31.3 (34.7) \\
\hline \hline
\end{tabular}
\caption{\label{tab:phippties} $\phi$ mass and width at  normal nuclear matter density $\rho_{0}$
  with and without the gauged Lagrangian of \eqn{eqn:phi2kk}. The values in parentheses are computed
  by adding  the gauged Lagrangian of \eqn{eqn:phi2kk}. All quantities are given in MeV.
For the cutoff values of $\Lambda_{K}=2000,3000$, and 4000 MeV we find  the $\phi$-meson bare mass 
$m^0_\phi$ values  
of 1074.0 (1023.4) MeV, 1132.9 (1024.6) MeV, and 1213.6 (1025.6) MeV, respectively.}
\end{center}
\end{table}

The results described above support a small downward mass shift and a large broadening of the decay
width of the $\phi$-meson in a nuclear medium. Furthermore, they open experimental possibilities for
studying the binding and absorption of $\phi$-mesons in nuclei.  Although the mass shift found in this
study may be large enough to bind the $\phi$-meson to a nucleus, the broadening of its decay width
will make it difficult to observe a signal for the $\phi$-nucleus bound state formation experimentally.
We explore this further in the following section.

\section{\label{sec:finitenuclei} $\phi$-nuclear bound states}

\begin{figure*}
  \begin{tabular}{c@{\hskip 5mm}c@{\hskip 5mm}c}
\includegraphics[scale=0.2,angle=-90]{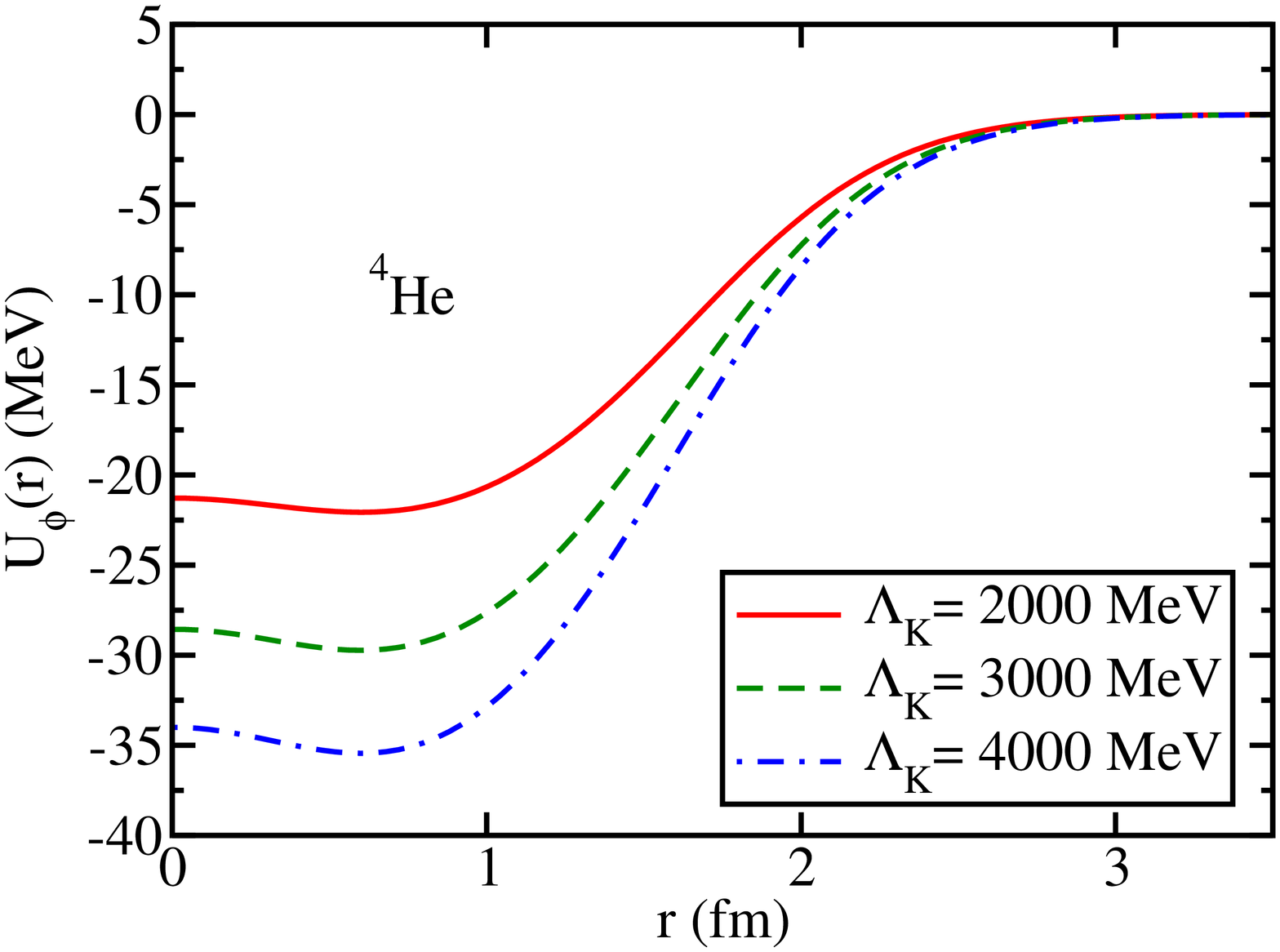} &  
\includegraphics[scale=0.2,angle=-90]{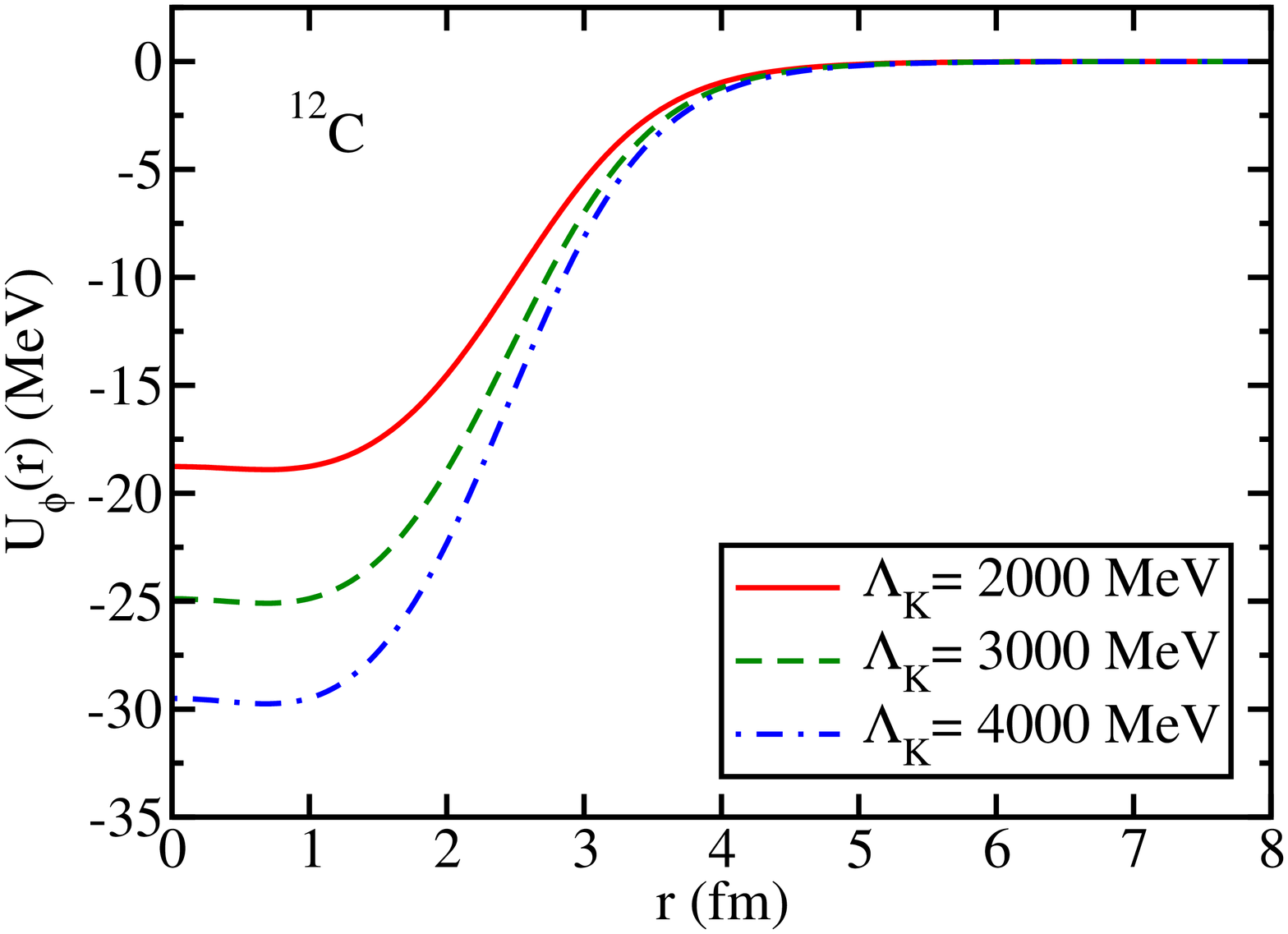} &
\includegraphics[scale=0.2,angle=-90]{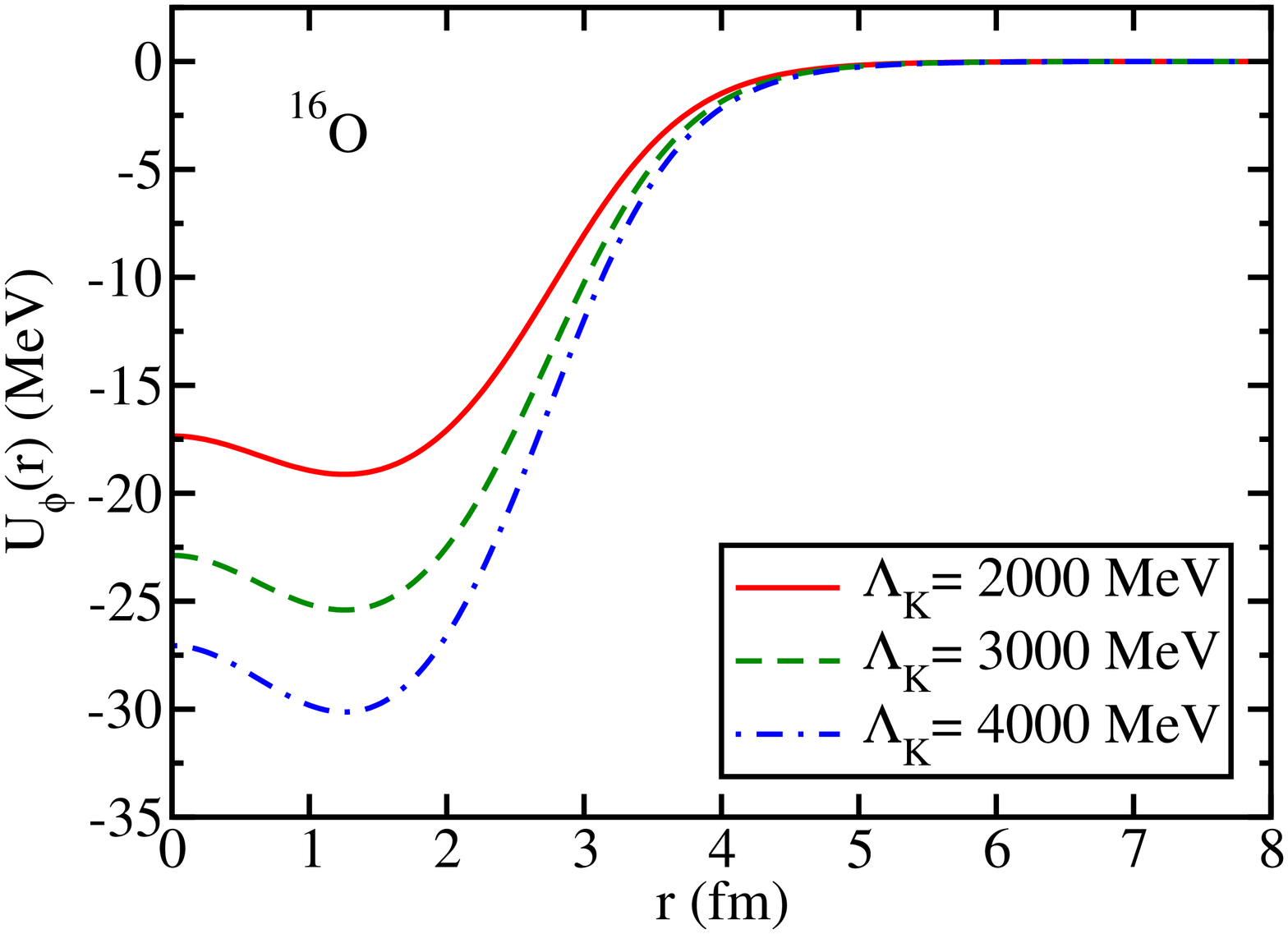} \\ [37mm]
\includegraphics[scale=0.2,angle=-90]{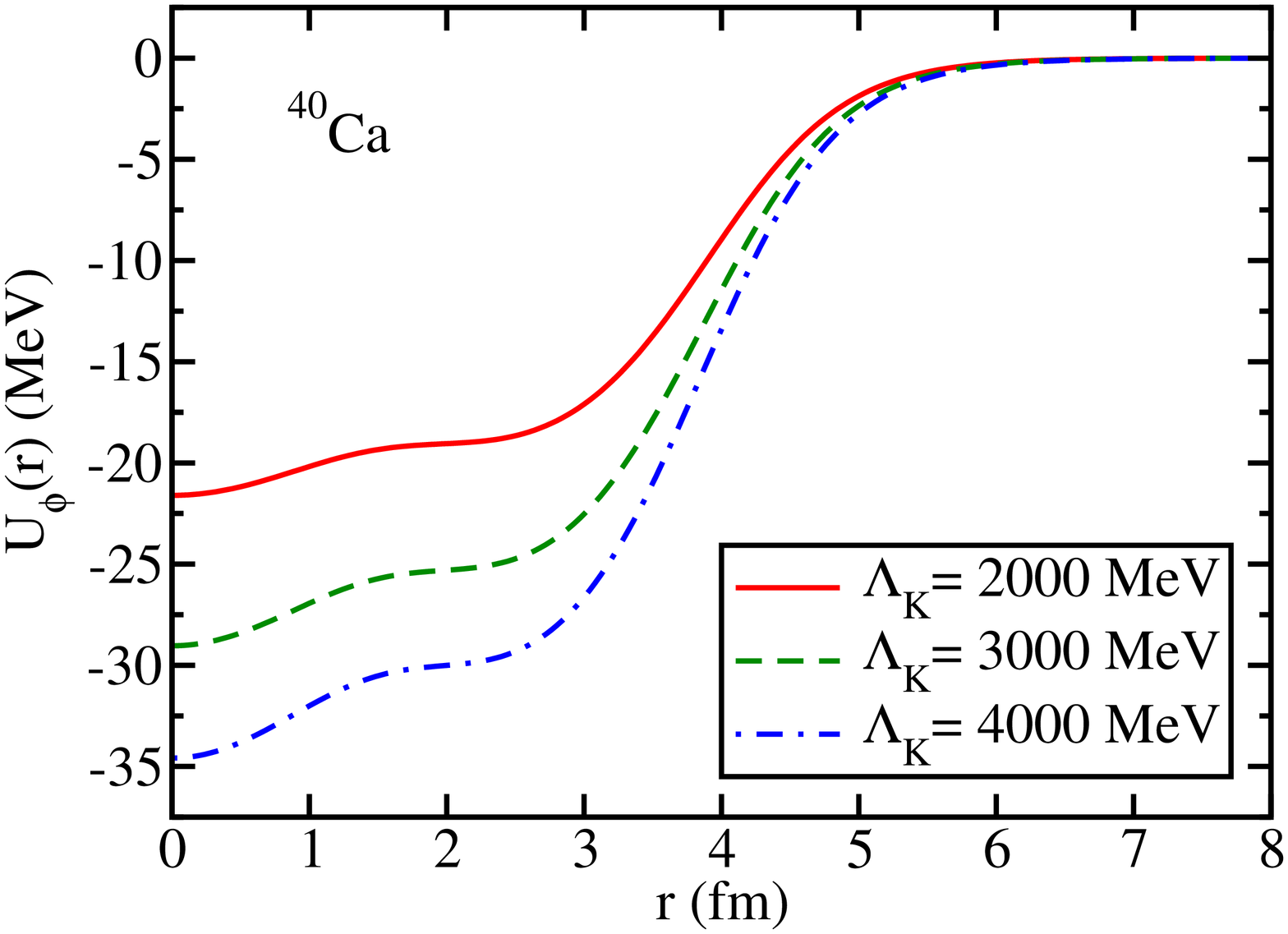} &
\includegraphics[scale=0.2,angle=-90]{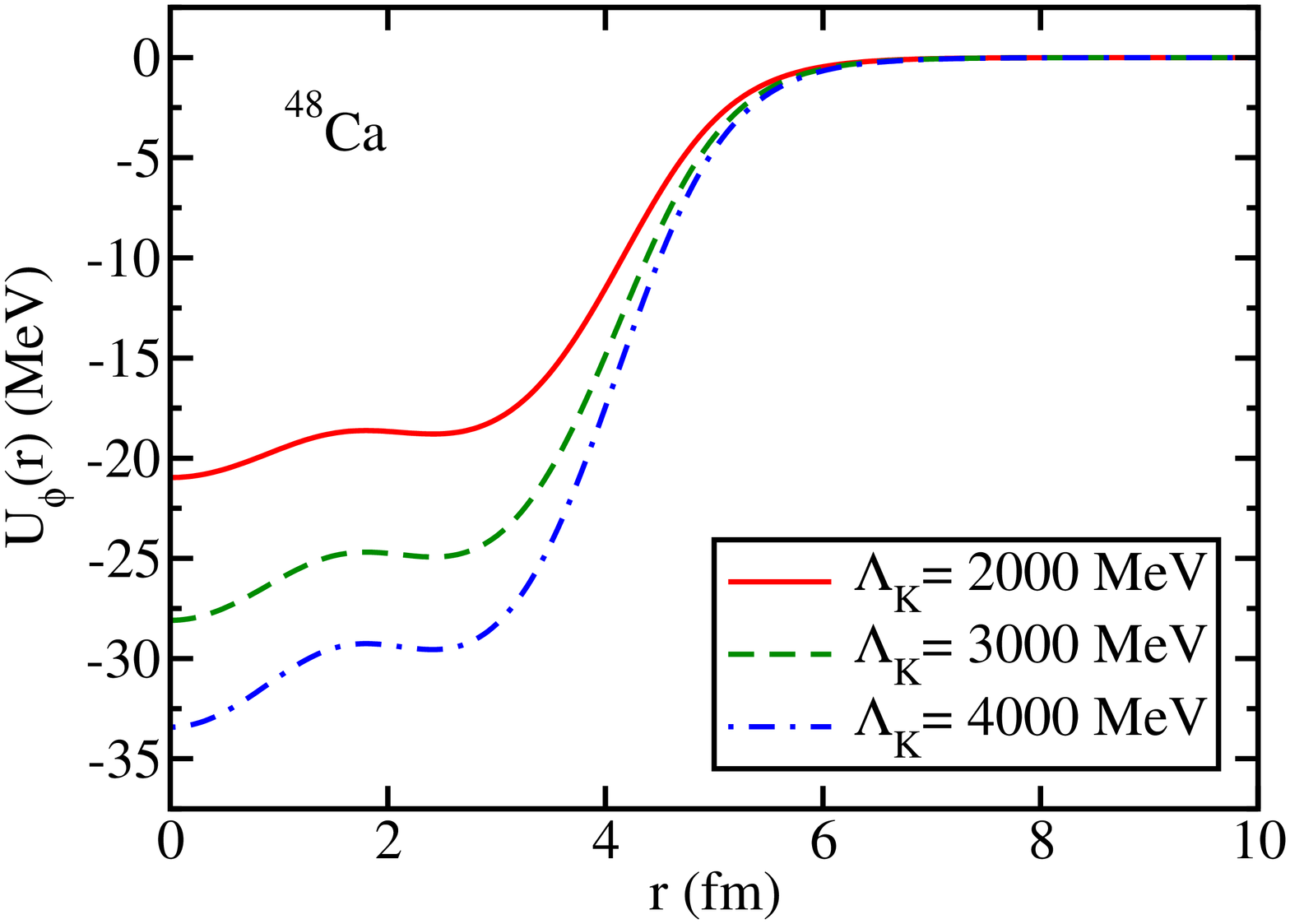} &
\includegraphics[scale=0.2,angle=-90]{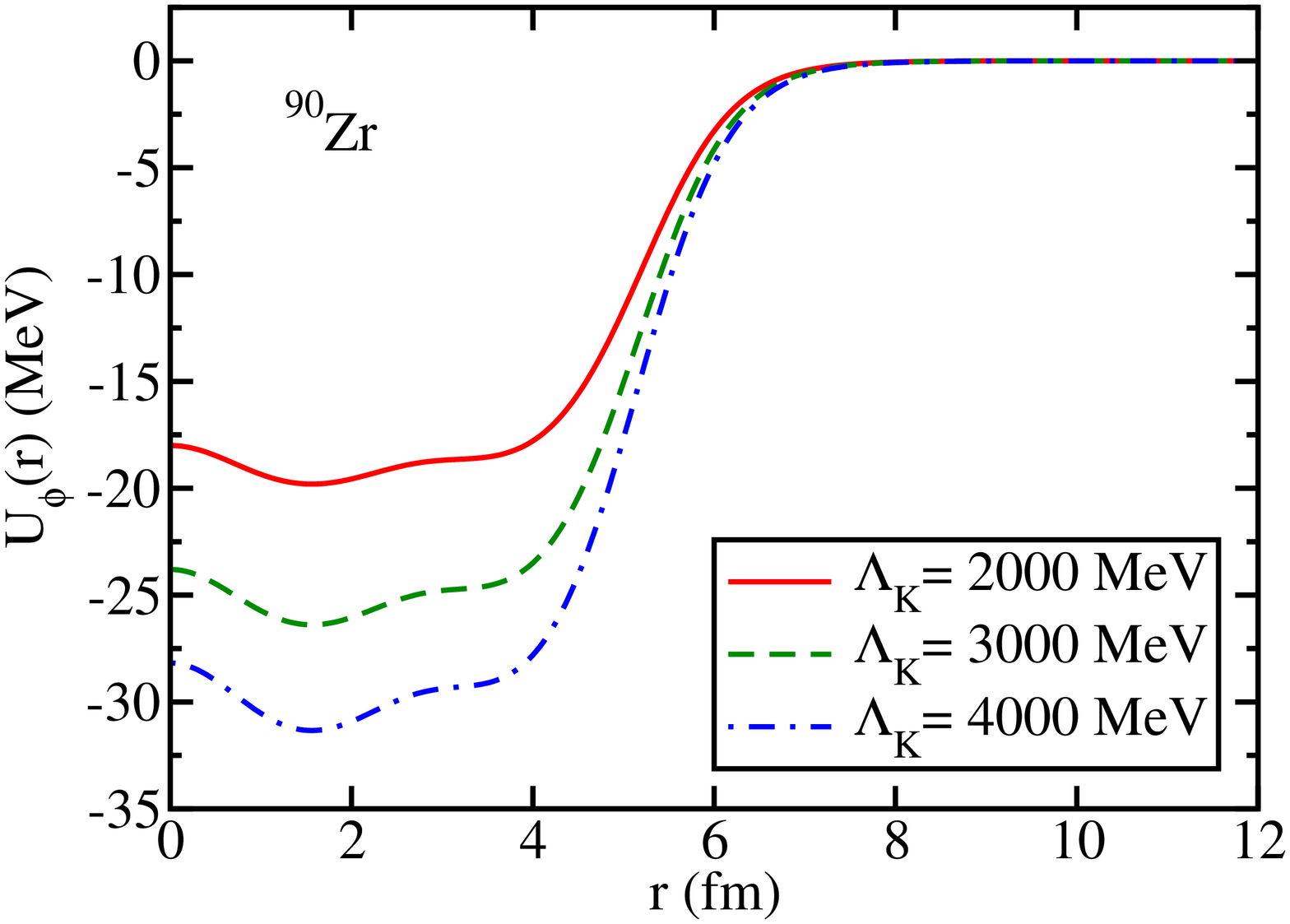} \\ [37mm]
\includegraphics[scale=0.2,angle=-90]{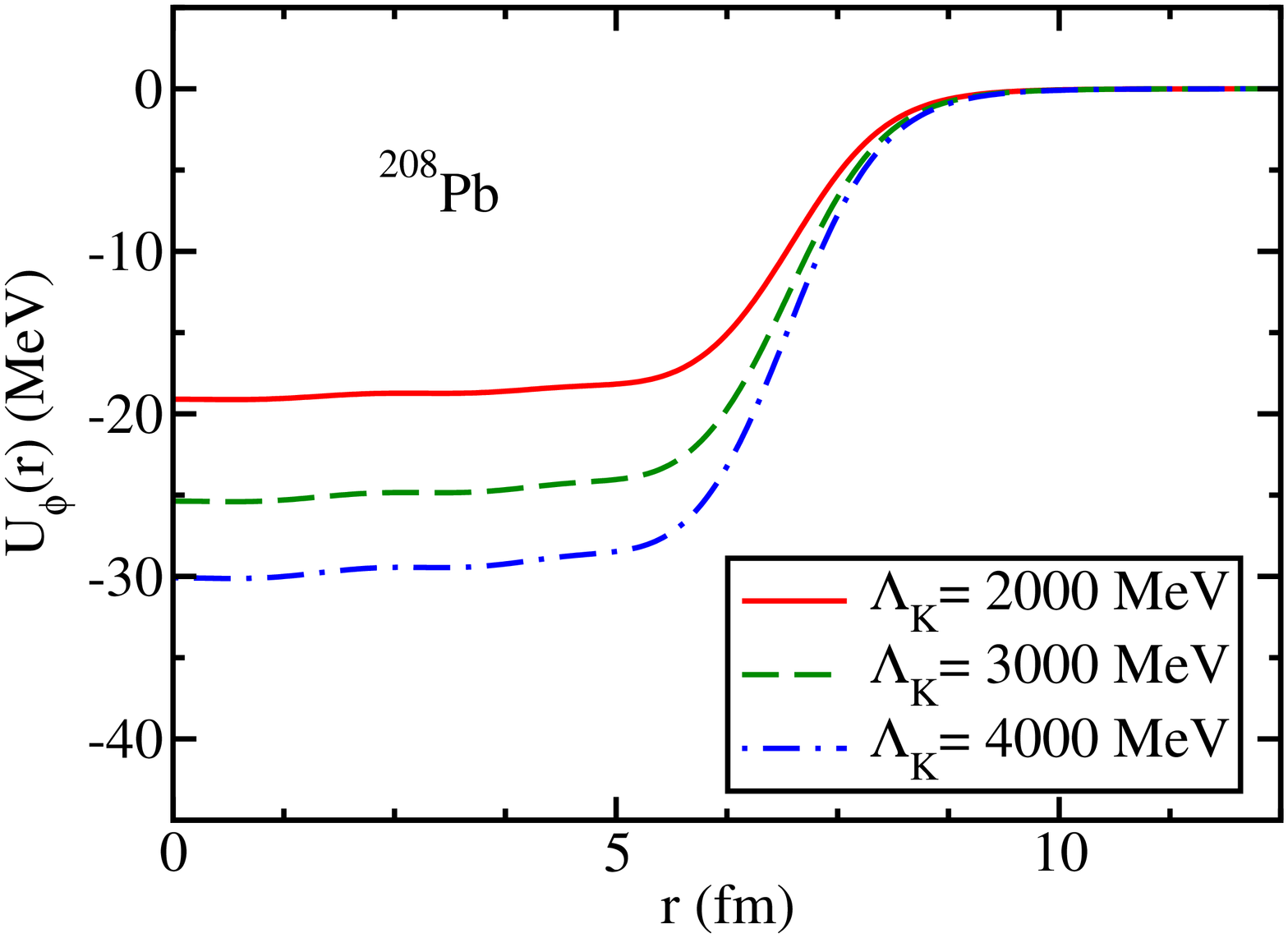} &
\includegraphics[scale=0.2,angle=-90]{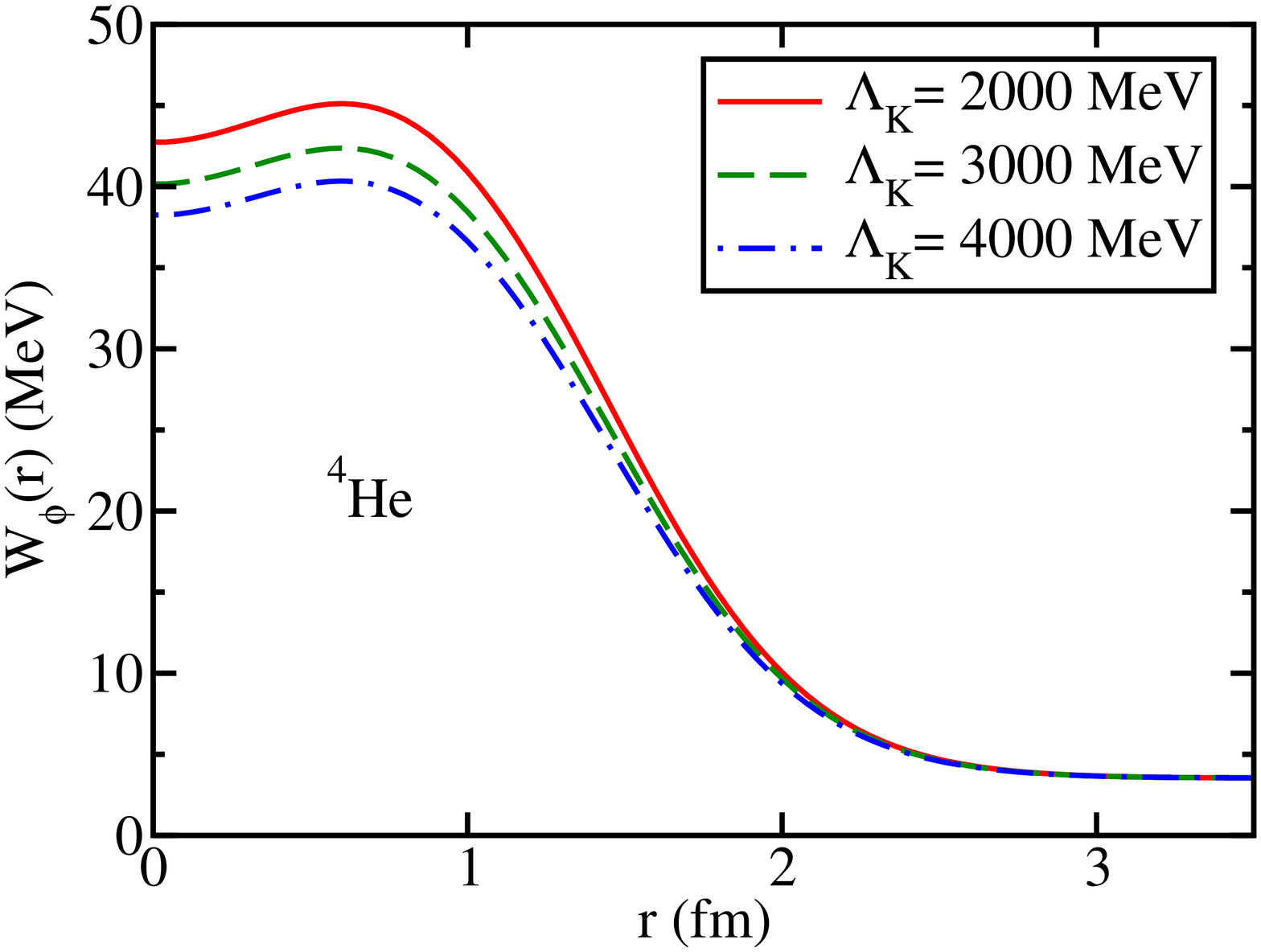} &
\includegraphics[scale=0.2,angle=-90]{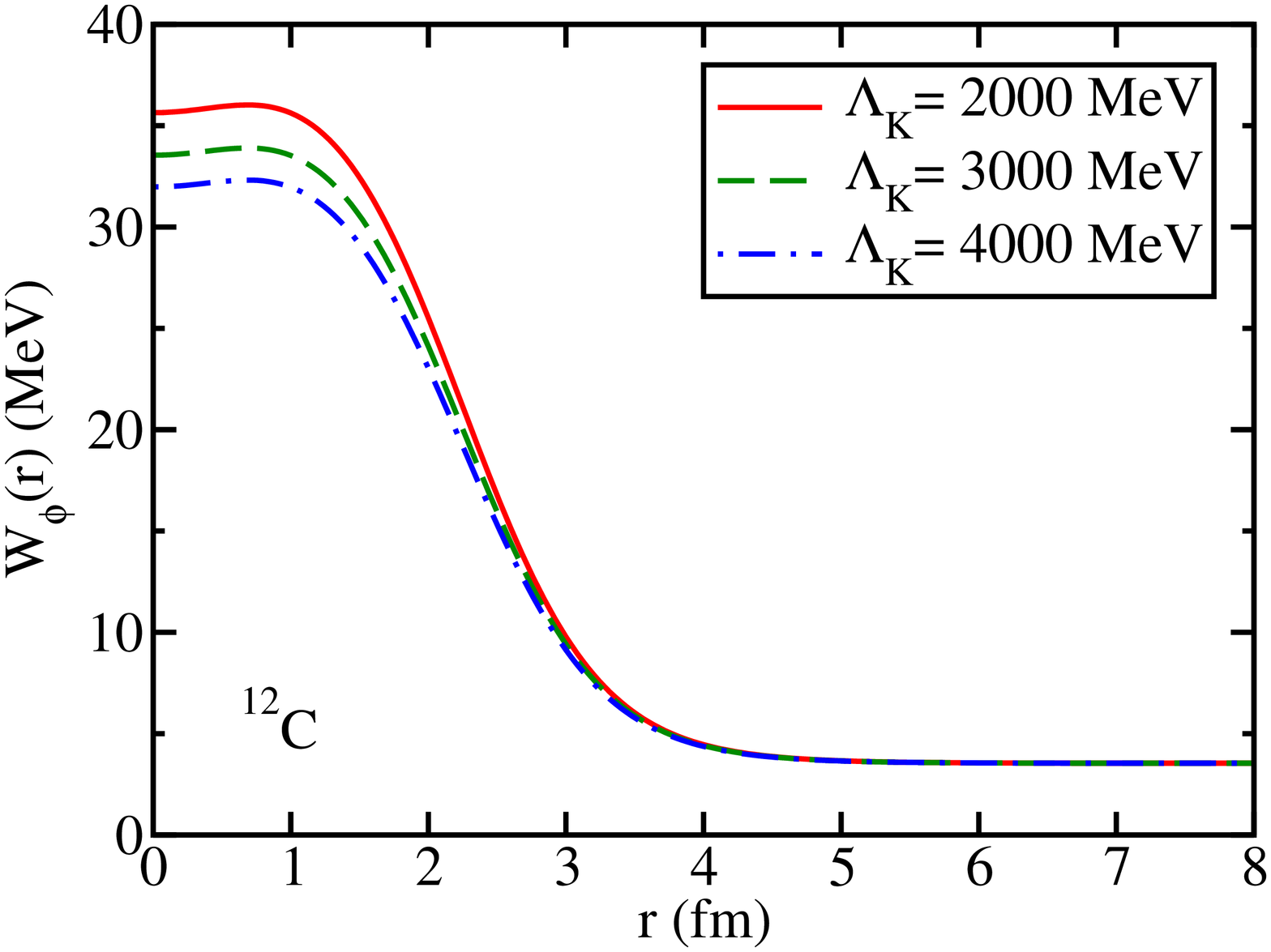} \\ [37mm]
\includegraphics[scale=0.2,angle=-90]{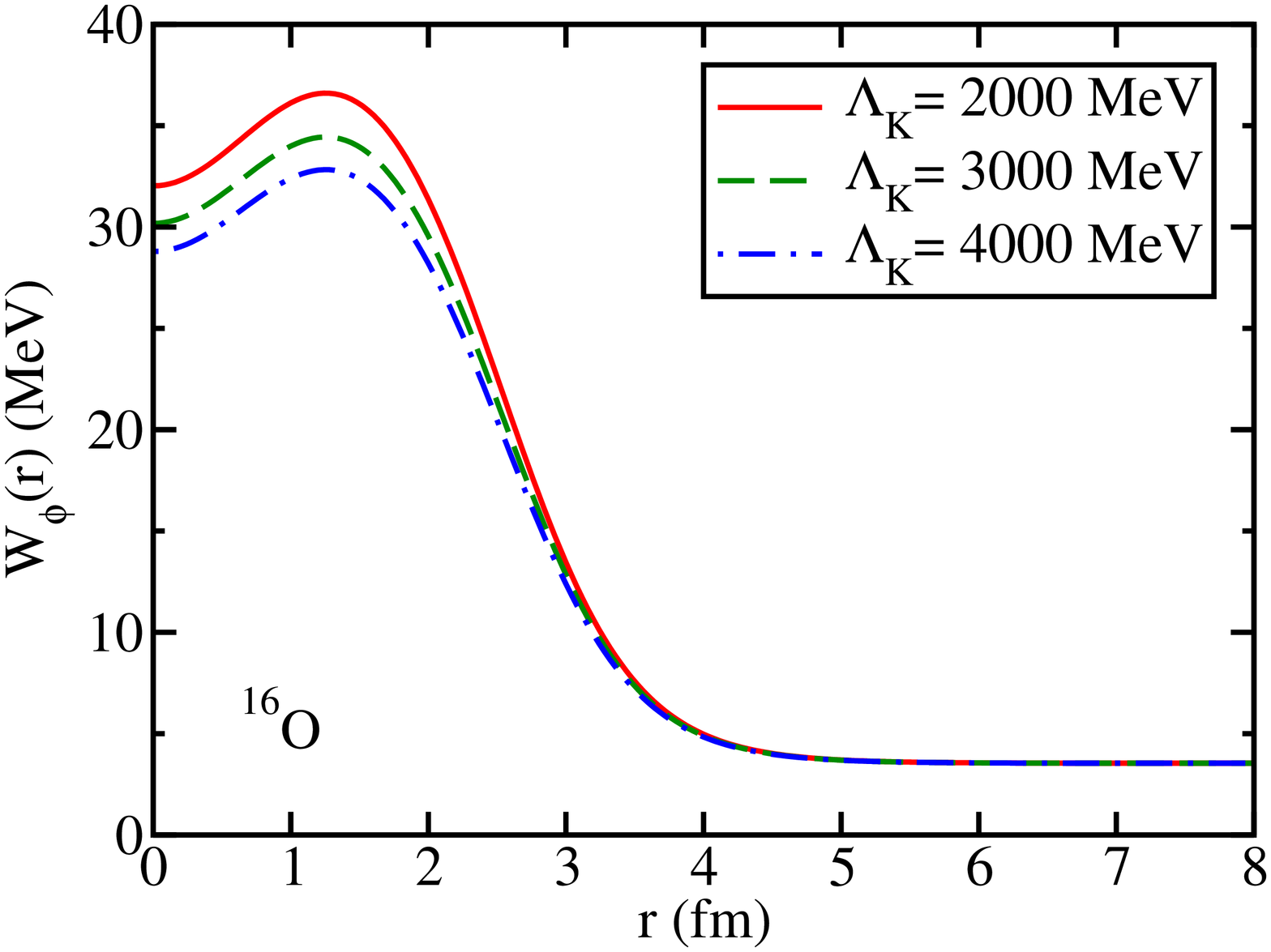} &
\includegraphics[scale=0.2,angle=-90]{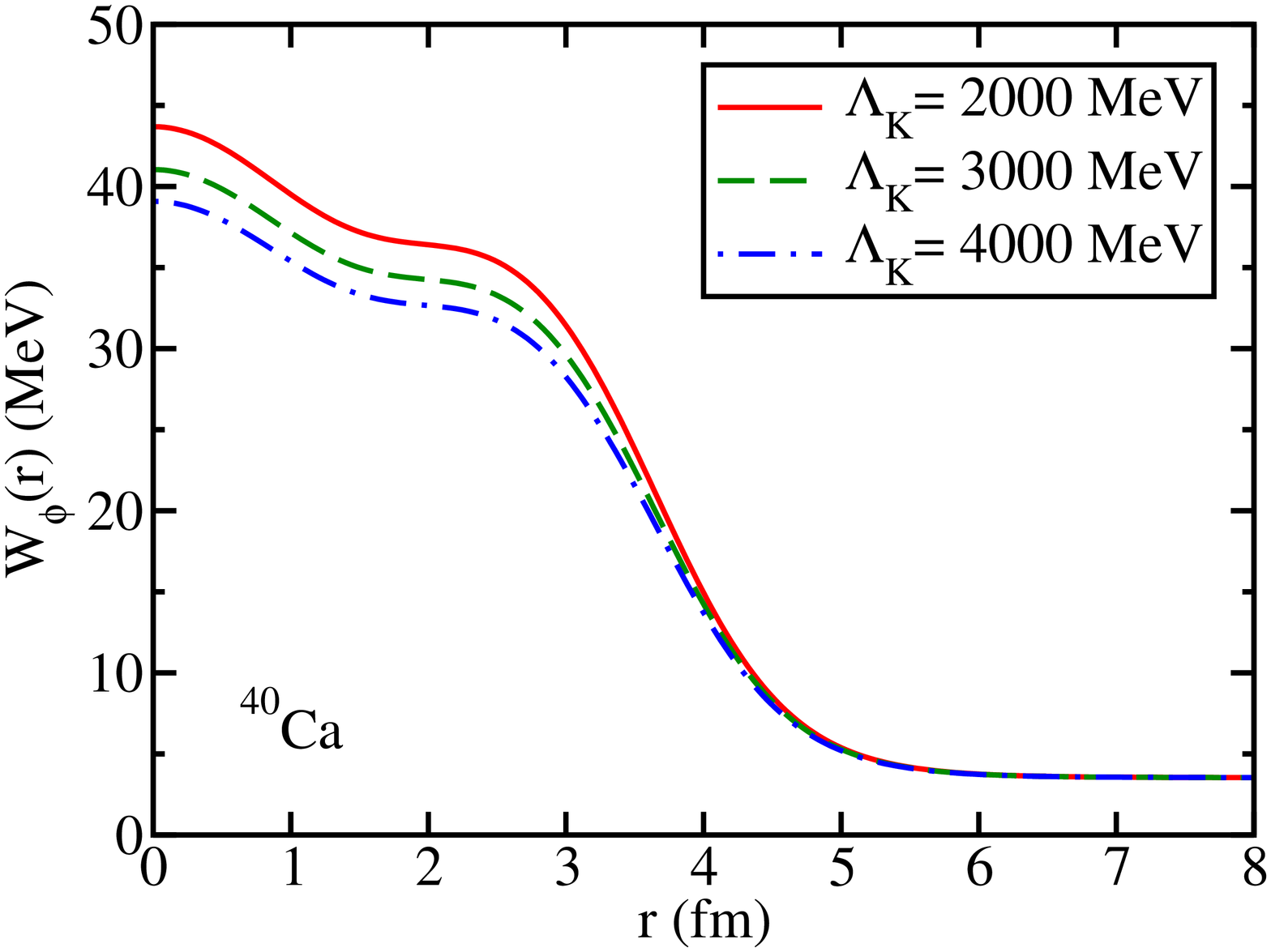} &
\includegraphics[scale=0.2,angle=-90]{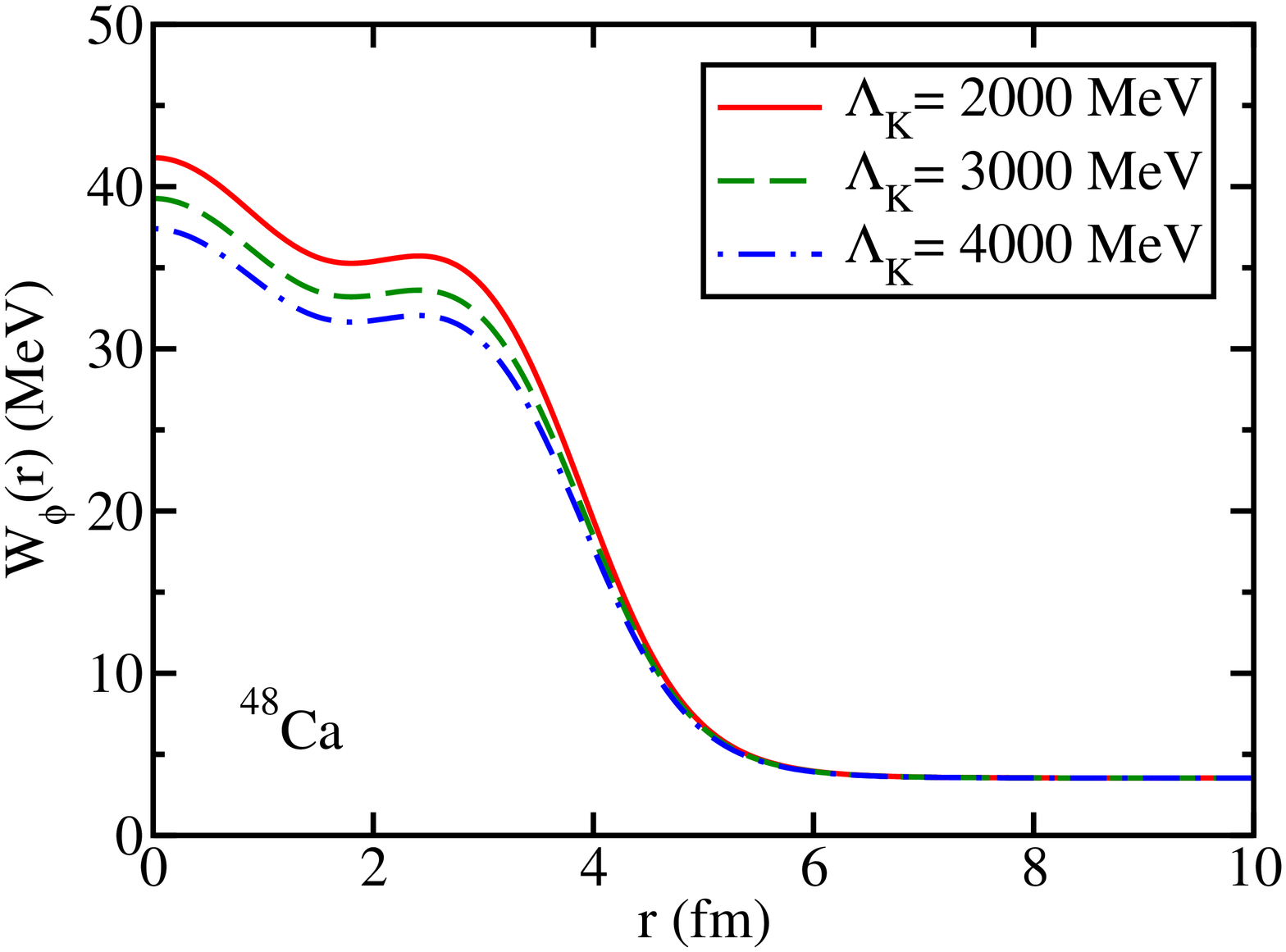} \\ [37mm]
\includegraphics[scale=0.2,angle=-90]{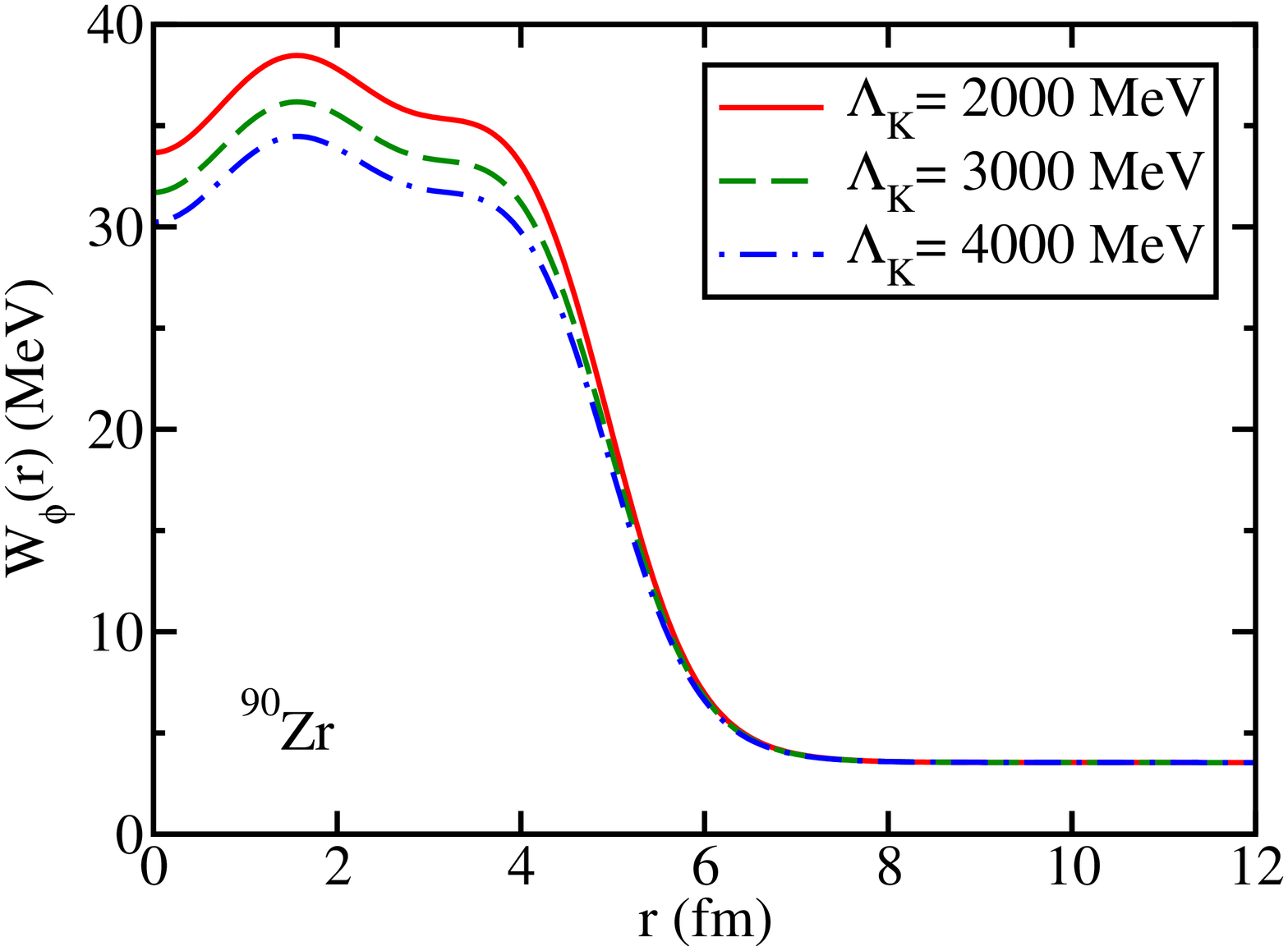} &
\includegraphics[scale=0.2,angle=-90]{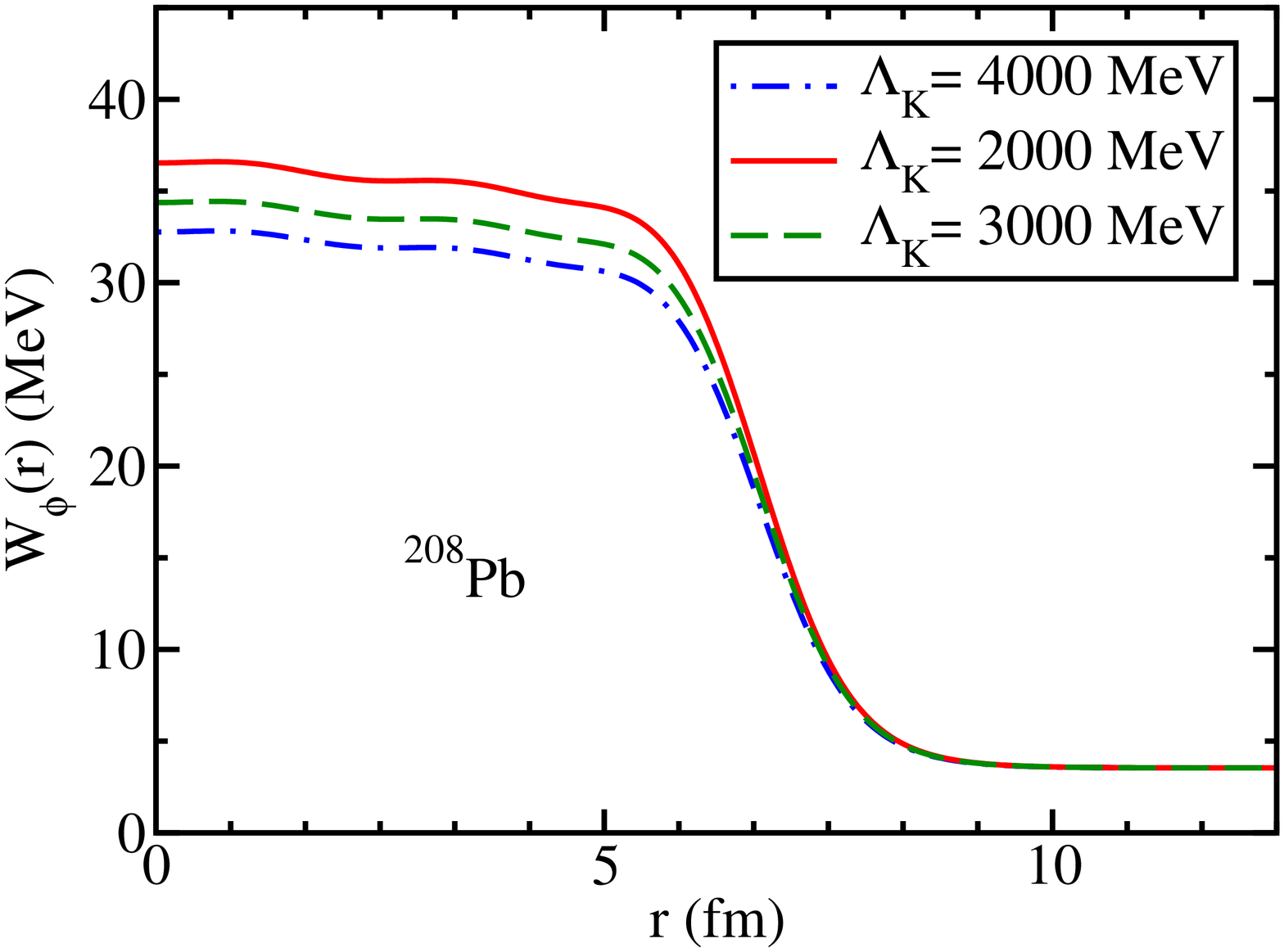} &
  \end{tabular}
  \caption{\label{fig:VphiA} Real [$U_{\phi}(r)$] and imaginary  [$W_{\phi}(r)$] parts of the
    $\phi$-meson-nucleus  potentials in seven nuclei selected,  for three values of the cutoff parameter
    $\Lambda_{K}$.}
\end{figure*}

In this section we discuss the situation where the $\phi$-meson is placed in  a nucleus. The nuclear density
distributions for $^{12}$C, $^{16}$O, $^{40}$Ca, $^{48}$Ca, $^{90}$Zr, and $^{208}$Pb  are obtained using
the QMC model~\cite{Saito:1996sf}. For $^{4}$He, we use the parametrization for the density distribution
obtained in Ref.~\cite{Saito:1997ae}. Then, using a local density approximation we calculate the
$\phi$-meson complex potentials for a nucleus $A$, which can be written as
\begin{equation}
\label{eqn:Vcomplex}
V_{\phi A}(r)= U_{\phi}(r)-\frac{\mi}{2}W_{\phi}(r),
\end{equation}
\noindent where $r$ is the distance from the center of the nucleus and
$U_{\phi}(r)=\Delta m_{\phi}(\rho_{B}(r)) \equiv m^{*}_{\phi}(\rho_{B}(r))-m_\phi$  and
$W_{\phi}(r)=\Gamma_{\phi}(\rho_{B}(r))$ are, respectively, the $\phi$-meson mass shift and decay width in
a nucleus $A$. As usual,  $\rho_{B}(r)$ is the baryon density distribution for the particular nucleus.

In Figure~\ref{fig:VphiA} we present the $\phi$-meson potentials calculated for the seven nuclei selected,
for three values of the cutoff parameter $\Lambda_{K}$, $2000, 3000$ and $4000$ MeV. 
One can see that the depth of the real part of the potential, $U_\phi(r)$, is sensitive to the cutoff
parameter, varying from -20 MeV to -35 MeV for $^{4}$He and from -20 MeV to -30 MeV for $^{208}$Pb. 
In addition, one can see that the imaginary part does not vary much with different values of $\Lambda_{K}$.
These observations may well have consequences for the feasibility of experimental observation of the
expected bound states.
%
\begin{table}[ht]
\begin{center}
\scalebox{0.85}{
  \begin{tabular}{ll|rr|rr|rr} 
\hline \hline
& & \multicolumn{2}{c|}{$\Lambda_{K}=2000$} &
\multicolumn{2}{c|}{$\Lambda_{K}=3000$} & 
\multicolumn{2}{c}{$\Lambda_{K}=4000$}  \\
\hline
 & & $E$ & $\Gamma/2$ & $E$ & $\Gamma/2$ & $E$ & $\Gamma/2$ \\
\hline
$^{4}_{\phi}\text{He}$ & 1s & n (-0.8) & n & n (-1.4) & n & -1.0 (-3.2) & 8.3 \\
\hline
$^{12}_{\phi}\text{C}$ & 1s & -2.1 (-4.2) & 10.6 & -6.4 (-7.7) & 11.1 & -9.8 (-10.7) & 11.2 \\
\hline
$^{16}_{\phi}\text{O}$ & 1s & -4.0 (-5.9) & 12.3 & -8.9 (-10.0) & 12.5 & -12.6 (-13.4) & 12.4 \\
& 1p & n (n) & n & n (n) & n & n (-1.5) & n \\
\hline
$^{40}_{\phi}\text{Ca}$ & 1s & -9.7 (-11.1) & 16.5 & -15.9 (-16.7) & 16.2 & -20.5 (-21.2) & 15.8 \\
& 1p & -1.0 (-3.5) & 12.9 & -6.3 (-7.8) & 13.3 & -10.4 (-11.4) & 13.3 \\
& 1d & n (n) & n & n (n) & n & n (-1.4) & n \\
\hline
$^{48}_{\phi}\text{Ca}$ & 1s & -10.5 (-11.6) & 16.5 & -16.5 (-17.2) & 16.0 & -21.1 (-21.6) & 15.6 \\
& 1p & -2.5 (-4.6) & 13.6 & -7.9 (-9.2) & 13.7 & -12.0 (-12.9) & 13.6 \\
& 1d & n (n) & n & n (-0.8) & n & -2.1 (-3.6) & 11.1 \\
\hline
$^{90}_{\phi}\text{Zr}$ & 1s & -12.9 (-13.6) & 17.1 & -19.0 (-19.5) & 16.4 & -23.6 (-24.0) & 15.8 \\
& 1p & -7.1 (-8.4) & 15.5 & -12.8 (-13.6) & 15.2 & -17.2 (-17.8) & 14.8 \\
& 1d & -0.2 (-2.5) & 13.4 & -5.6 (-6.9) & 13.5 & -9.7 (-10.6) & 13.4 \\
& 2s & n (-1.4) & n & -3.4 (-5.1) & 12.6 & -7.4 (-8.5) & 12.7 \\
& 2p & n (n) & n & n (n) & n & n (-1.1) & n \\
\hline
$^{208}_{\phi}\text{Pb}$ & 1s & -15.0 (-15.5) & 17.4 & -21.1 (-21.4) & 16.6 & -25.8 (-26.0) & 16.0 \\
& 1p & -11.4 (-12.1) & 16.7 & -17.4 (-17.8) & 16.0 & -21.9  (-22.2) & 15.5 \\
& 1d & -6.9 (-8.1) & 15.7 & -12.7 (-13.4) & 15.2 & -17.1 (-17.6) & 14.8 \\
& 2s & -5.2 (-6.6) & 15.1 & -10.9 (-11.7) & 14.8 & -15.2 (-15.8) & 14.5 \\
& 2p & n (-1.9) & n & -4.8 (-6.1) & 13.5 & -8.9 (-9.8) & 13.4 \\
& 2d & n (n) & n & n (-0.7) & n & -2.2 (-3.7) & 11.9 \\
\hline \hline
\end{tabular}}
\caption{\label{tab:phibse} $\phi$-nucleus single-particle energies, $E$, and half widths, $\Gamma/2$,
  obtained with and without the imaginary part of the potential, for three values of the cutoff parameter
  $\Lambda_K$. When only the real part is included, where the corresponding single-particle energy $E$
  is given inside brackets, $\Gamma=0$ for all nuclei.``n'' indicates that no bound state is found.
All quantities are given in MeV.}
\end{center}
\end{table}

Using the $\phi$-meson potentials obtained in this manner, we next calculate the $\phi$-meson--nuclear
bound state energies and absorption widths for the seven nuclei selected. Before proceeding, a few
comments on the use of \eqn{eqn:kg} are in order.  In this study we consider the situation where the
$\phi$-meson is produced nearly at rest. Then, it should be a very good approximation to neglect the
possible energy difference between the longitudinal and transverse components of the $\phi$-meson
wave function $\psi_{\phi}^{\mu}$. After imposing the Lorentz condition, $\partial_{\mu}\psi_{\phi}^{\mu}=0$,
to solve the Proca equation becomes equivalent to solving the Klein-Gordon equation
\begin{equation}
\label{eqn:kg}
\left(-\nabla^{2} + \mu^{2} + 2\mu V(\vec{r})\right)\phi(\vec{r})
= \mathcal{E}^{2}\phi(\vec{r}),
\end{equation}
\noindent where $\mu=m_{\phi}m_{A}/(m_{\phi}+m_{A})$ is the reduced mass of the $\phi$-meson-nucleus
system with $m_{\phi}$ $(m_{A}$) the mass of the $\phi$-meson (nucleus $A$) in vacuum, and $V(\vec{r})$
is the complex $\phi$-meson-nucleus potential of \eqn{eqn:Vcomplex}. We solve the Klein-Gordon
equation using the momentum space methods developed in Ref.~\cite{Kwan:1978zh}. Here, \eqn{eqn:kg}
is first converted to momentum space representation via a Fourier transform, followed by a partial
wave-decomposition of the Fourier-transformed potential. Then, for a given value of angular momentum,
the eigenvalues of the resulting equation are found by the inverse iteration eigenvalue algorithm.
The calculated bound state energies ($E$) and widths ($\Gamma$), which are related to the complex
energy eigenvalue $\mathcal{E}$ by $E=\Re\mathcal{E}-\mu$ and $\Gamma=-2\Im\mathcal{E}$, are listed
in \tab{tab:phibse} for three values of the cutoff parameter $\Lambda_{K}$, with and without the imaginary
part of the potential, $W_{\phi}(r)$.

We first discuss the case in which the imaginary part of the $\phi$-nucleus potential, $W_{\phi}(r)$, is set to
zero.  The results are listed inside brackets in \tab{tab:phibse}. From the values shown in brackets, we see
that the $\phi$-meson is expected to form bound states with all the seven nuclei selected, for all values of
the cutoff parameter $\Lambda_{K} = 2000, 3000$ and $4000$ MeV. (For the variation in the potential
depths due to the $\Lambda_K$ values, see Fig.~\ref{fig:VphiA}.) However, the bound state energy is
obviously dependent on $\Lambda_{K}$, increasing as $\Lambda_K$ increases.

Next, we discuss the results obtained when the imaginary part of the potential is included. Adding the
absorptive part of the potential,  the situation changes appreciably. From the results presented in
\tab{tab:phibse} we note that for the largest value of the cutoff parameter $\Lambda_K = 4000$ MeV, 
which yields the deepest attractive potentials, the $\phi$-meson is expected to form bound states in all
the nuclei selected, including the lightest $^4$He nucleus. However, in this case, whether or not the bound
states can be observed experimentally is sensitive to the value of the cutoff parameter $\Lambda_K$.
One also observes that the width of the bound state is insensitive to the values of $\Lambda_{K}$ for all
nuclei. Furthermore, since the so-called dispersive effect of the absorptive potentialis repulsive, 
the bound states disappear completely in some cases, even though they were found when the absorptive
part was set to zero. This feature is obvious for the  $^4$He nucleus, making it especially relevant to the
future experiments, planned at J-PARC and JLab using light and medium-heavy nuclei~\cite{Buhler:2010zz,
Ohnishi:2014xla,Csorgo:2014sat,JLabphi}.

We here comment that we have also solved the Schr\"{o}edinger equation with the potential
\eqn{eqn:Vcomplex} with and without its imaginary part for the single-particle energies and widths, and
compared with those given in \tab{tab:phibse}. The results found in both cases are essentially the same.

\section{\label{sec:summary} Summary and discussion}

We have calculated the $\phi$-meson--nucleus bound state energies and absorption widths for various
nuclei. The $\phi$-meson--nuclear potentials  were calculated using a local density approximation, with
the inclusion of the $K\Kbar$ meson loop in the $\phi$-meson self-energy. The nuclear density
distributions, as well as the in-medium $K$ and $\Kbar$ meson masses, were consistently calculated by
employing the quark-meson coupling model. Using the $\phi$-meson--nuclear complex potentials, we have
solved the Klein-Gordon equation in momentum space,  and obtained $\phi$-meson--nucleus bound state
energies and absorption widths. Furthermore, we have studied the sensitivity of the results to the cutoff
parameter $\Lambda_{K}$ in the form factor at the $\phi-K\Kbar$ vertex appearing in the $\phi$-meson
self-energy. We expect that the $\phi$-meson should form bound states for all seven nuclei selected,
provided that the $\phi$-meson is produced in (nearly) recoilless kinematics. This feature, is even more
obvious in the (artificial) case where the absorptive part of the potential is ignored. Given the similarity
of the binding energies and widths reported here, the signal for the formation of the $\phi$-nucleus bound
states  may be difficult to identify experimentally.  Therefore, the feasibility of observation of the
$\phi$-meson--nucleus bound states needs further investigation, including explicit reaction cross section
estimates. \\

\section*{Acknowledgements}
This work was partially supported by Conselho Nacional de Desenvolvimento Cient\'{i}fico e
Tecnol\'{o}gico - CNPq, Grant Nos. 152348/2016-6 (J.J.C-M.), 400826/2014-3 and 308088/2015-8 (K.T.),
305894/2009-9 (G.K.), and 313800/2014-6 (A.W.T.), and Funda{\c c}\~{a}o de Amparo \`{a} Pesquisa do
Estado de S\~{a}o Paulo-FAPESP, Grant Nos. 2015/17234-0 (K.T.) and 2013/01907-0 (G.K.).
This research was also supported by the University of Adelaide and by the Australian Research Council
through the ARC Centre of Excellence for Particle Physics at the Terascale (CE110001104), and through
Grant No. DP151103101 (A.W.T.).



\end{document}